# Holistic Approach for Critical System Security: Flooding Prevention and Malicious Packet Stopping

Mohammed A. Alhabeeb, Abdullah Almuhaideb, and Phu Dung Le

**Abstract**— Denial of service attacks (DoS) can cause significant financial damages. Flooding and Malicious packets are two kinds of DoS attacks. This paper presents a new security approach which stops malicious packets and prevents flooding in the critical systems. New concepts of packet stamp a dynamic-multi-communication-point mechanism has been identified for this proposed approach to make the prevention of flooding attacks easier and the performing of malicious packet attacks harder. In addition, dynamic key encryption technique has been adapted as a part of the proposed approach to enhance its functionality.

——————————— ◆ ———————————

## 1 INTRODUCTION

Technologies in the information age provide vast opportunities for organizations to transform their services into the digital arena. For some organizations, like government departments, which have very critical systems, security is one of the important factors to provide online services. Information availability is one of the important security goals. It means that providing services should be uninterrupted by malicious DoS [1], [2], [3], [7]. According to Kim and Kim in [18], communication privacy is one of the importance principals for a critical system to provide online services, so a strong encryption mechanism must be used to achieve this goal. The Internet has been designed to maximise its functionality to provide communication, and its security was not considered to be a major factor [6]. Recently, DoS attacks against highly visible Internet sites or services have become commonplace [4]. So attacks have been a danger to the Internet operations, and they have caused significant financial damages [3][5]. According to the 2009 CSI Computer Crime and Security Survey report, 29.2% of the respondents detected DoS attacks directed against them, with the respondents representing that the DoS attacks were the most costly cyber attack for them [8].

DoS caused by IP packet floods is one of the major problems faced by Internet hosts. It is nearly impossible to stop packets addressed to the host. IP routers respond to dropping packets arbitrarily when an overload happens. But the question is which packet should be dropped [9]. So, flooding attacks are hard to detect [10]. It can effortlessly degrade the Quality of Service (QoS) in the network and leads to the interruption of critical infrastructure services [15]. Flooding attacks are a serious threat to the security of networks that provide public services like government portals [16]. They are more difficult to fight against if the IP has been spoofed [17]. Flooding attacks are easier to be committed when the encryption data is included in the security solutions, because an addition of overload process for communication comes from encryption and decryption of the data [18].

Malicious packet attack is a type of the DoS attacks. It also called malformed packet attack. It occurs when the attacker sends incorrectly formatted packets to the victim system to crash it [30].There are two types of malicious packet attacks: packet address attacks and packet attribute attacks [30]. These kinds of attacks are continuously growing, because the attackers identify the limitation of protocols and applications from time to time [31]. Some solutions have been developed to defence against malicious packet attacks. However attackers are continuously searching for system vulnerabilities to commit new malicious packet attacks [32]. Hackers today are trying to find systems vulnerabilities by generating random packets, which have different attribute possibilities [32]. Old malicious packet attacks can also still be found [31]. One reason for that is because some of the new operating system and devices are vulnerable for some of the old attacks. For example, the beta version of Windows Vista was vulnerable to a number of old vulnerabilities [32].

In this paper, a new security approach, which is Holistic Approach for Critical System Security (HACSS), is proposed. It is designed for critical systems like government portals. In these systems, higher availability of services and higher privacy of communication are important principles. The HACSS is identified to support these systems' availability by preventing DoS attacks like flooding and malicious packet attacks, and to enable a suitable encryption solution. In this paper we will illustrate how the HACSS deals with flooding and malicious packet attacks in these systems.

————————————————

- M.A. Alhabeeb is with the Faculty of Information Technology, Monash University , Melbourne, Australia.
- A.M Almuhaideb is with the Faculty of Information Technology, Monash University , Melbourne, Australia.
- P.D Le is with the Faculty of Information Technology, Monash University , Melbourne, Australia.

Many of DoS attacks come from packets. In the HACSS, the connection is designed to be controlled by the server, thus the client's packets are designed by the server. In addition, this approach divides the communications with clients into two groups: communication with authenticated clients and communication with non-authenticated clients. These two groups communicate with the system via two separate channels. This division helps the system to expect what kinds of packets should be received in each part, and it also helps the system to design the suitable solution to stop malicious packet attacks and to prevent flooding attacks. So this approach should stop all existing malicious packet attacks, and it will prevent a lot of future malicious packet attacks. The HACSS identifies a new concept which we call packet stamp technique. This technique is designed to stop malicious packet attacks from authenticated clients. So the system can provide services like uploading files or receiving messages while stopping malicious packet attacks. In addition, the HACSS enables a new dynamic-multi-points-communication mechanism, which makes the prevention of flooding attacks easier, and it makes it easier to stop spoofed malicious packet attacks. In addition, this approach includes dynamic key encryption to prevent sniffing, which leads to flooding attacks and reuse packet attacks which is kind of malicious packet attacks. These kinds of attacks occur when the attacker sniffs an authenticated client's packets, and then resends a packet many times to the server with slightly changed attributes.

The next section will further describe flooding and malicious packet attacks and will also discuss related works as security against these kinds of attacks. Then the proposed approach and its components will be explained in Section 3. The last section, will discuss how this HACSS will prevent flooding attacks and how stopping malicious packet attacks, and will also evaluate the HACSS in preventing flooding and in stopping malicious packets.

## 2 BACKGROUND

DoS attacks are committed by using victim resources to slow down or stop a key resource (CPU, bandwidth, buffer, etc) or more of these victim resources. The goal of DoS is either slowing down or stopping victim resources from providing services to the clients [6]. In the following two subsections, we will descuss how both of these attacks can be committed. In addition we will illustrate some of the previous solution against each of them.

### 2.1 Flooding attack

Flooding is one kind of DoS attacks. It occurs when an attacker sends an overpowering quantity of packets to a victim site. Some victim key resources might be crashed or delayed from responding as a result of handling this quantity of packets [6] as shown in (Fig. 1) below. Encryption of communications is a significant feature that must be included in critical system security designed to achieve confidentiality [14]. However, strong encryption might lead to flooding, because the system needs more time to process an encrypted packet that might be dropped. Dropping flooding packets from their headers is the easiest solution.

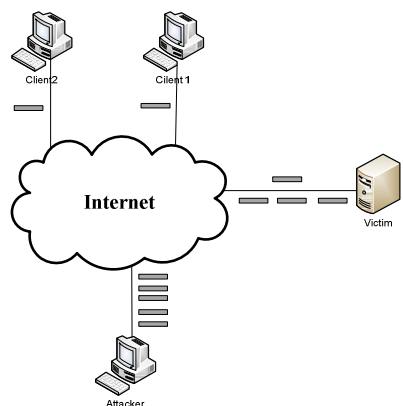

**Fig. 1.** Flooding attack

There are some methods designed to prevent flooding attacks. One of the existed techniques is an active queue management policy. As a queue management policy is applied on to the server, the size of a backlog queue increases depending on the availability of the system resources, and the timeout period would decrease. So the number of free slots will increase in the backlog queues for new requests. Random Drop (RD) is another method to solve the problem of flooding attacks. It randomly replaces a new request with a request which was stored in the backlog queue when the backlog queue is full [19].

Both of these methods are designed to receive new requests after flooding is happened or about to happen. The active queue management policy solution expands the availability of the queue, to handle new requests depending on the system resources. However, if the system resources are limited, or if the queue was already expanded with no free slots, the system will still face a potential risk of flooding attacks. Also in RD solution, there is a risk that right requests might be replaced with attackers' requests. So the flooding attack problem is still unsolved, and both the solutions are based on the system resources capabilities. Neither of them makes any change in the system design or process to handle more requests by using the same existing resources. With these issues, it is necessary for any future solution to consider the ways to save system resources availability to handle more flooding attacks requests.

### 2.2 malicious packet attack

The attacker might use the victim's system vulnerabilities to commit DoS attacks [6]. Network protocol vulnerabilities are identified repeatedly from time to time. The system might be vulnerable to receive a malicious packet, which might crash its protocol. The attacker creates this malicious packet by changing packet's attributes, so the receiver protocol fails to handle this packet and loses its consistency [20]. All protocols are softwares which are written using programming languages. As in any programs, they might have vulnerabilities which can be influenced by some malicious inputs [21]. Usually these kinds of products have been tested strongly for potential vulnerabilities. However, some input vulnerabilities are often found in these protocols after they were used. More new attacks might be discovered in the future, as these new attacks are uncovered by these programs. In addition, some protocol vulnerabilities come from the specific nature of the languages with which the protocols have been written [22].

Malicious packets are also a threat to the Internet, as various software vulnerabilities allow attackers to achieve remote control of routers in the Internet [23]. Malicious packet attacks could be more powerful when it is used with other attacking techniques, like shorter distance fraud. This kind of attacks might disrupt network communication completely. Existing network protocols might not be able to detect malicious packet attacks [23].

Most of existing solutions against malicious packet attacks depend on filtering the malicious packets and dropping them [29], [26], [27], [28]. The difference between these solutions is in the method of finding malicious packets. One category of these methods is based on the analysis of packet headers using intelligent algorithms to detect malicious packets. Decision Tree is an example for this category. It is simple data mining method, which is designed like a tree which consists of branch nodes. Each node represents a choice from a number of alternatives, and every leaf node represents a class of data. It has learning algorithms to take a decision about packets. For malicious packet attacks, the learning data contains every encoded data of malicious packet attacks [25]. The second category is simply designed to find malicious packets using some rules, which describe likely attributes of malicious packets. Deep Network Packet Filtering is one example for this category. It works as multi layered filtering according to some rules based on the past known attacks that. It could be implemented in the firewall, or as a special equipment, such as a intrusion detection system [24].

The above solutions are essentially based on the techniques to find and drop potential malicious packet attacks, and they are open to receive any malicious packet. For the first category of these techniques, must be enough powerful for existing malicious packet attacks, also they were designed to find future packet attacks. However, these types of solutions do not guarantee to stop new packet attacks, because some of the future packet attacks may not be detected by their intelligent algorithms. The second category only deals with the existing malicious packet attacks. So, some of the future packet attacks might be not cached by these methods.

## 3 HOLISTIC APPROACH FOR CRITICAL SYSTEM SECURITY (HACSS)

In this section, an overview of the HACSS will be illustrated. In addition, each component of the HACSS will be described, and communication of these components in each of them will also be explained.

### 3.1 An overview of HACSS

As in the figure below (Fig. 2), our approach consists of two main components, Client Authentication (CA) and Authenticated Client Communication (ACC). Each one of these components is responsible to communicate with the client, depending on the stage in which the client communicates with the system. This division is important to identify the nature of clients' communication activities with the system, and this helps to give an appropriate powerful solution for flooding attacks in every part of the system. In addition, this division helps to give an appropriate powerful solution for malicious packet attacks in every part of the system also. So the system can provide securely all required communication services. Each communication service is provided through a specific part of the system.

Client authenticity and services providence for clients will be determined in the first component. Filtering and Redirect Engine is the first element in this component. It receives clients' requests, filters them, and accepts only correct requests. The second element in the CA component is Ticket Engine. It authenticates clients and issues different categories of tickets for them. A ticket's category is determined based on the services that the client can be provided with.

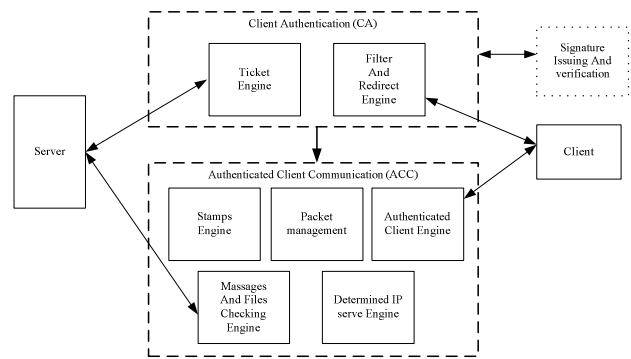

Fig. 2. HACSS Approach

Depending on the client satisfaction in the CA component, all communications for authenticated clients are moved to the ACC component. In this component, all clients' packets will be examined, and whole massages will be checked. It consists of five main elements, which are Authenticated Client Engine, Packet Manager, Stamp Engine, Determine IP Serve Engine, and Massage and File Checking Engine. The first four elements work together to accept only authenticated clients' packets, to design clients' packets, and to examine received packets. The Massage and File Checking Engine element checks all received clients' messages.

### 3.2 Client Authentication

The CA component handles all unauthenticated clients' requests (Fig. 3). It filters clients' requests, authenticates clients who hold these requests, and issues tickets for each client depending on the services that each client can receive. Only correct requests are accepted to be processed in the system. The CA provides a secure channel to the client while it is authenticating the client. Each client is required to provide his/her signature to represent his/her identity. All invalid signatures are added to a list. After the client is authenticated, a ticket will be issued for the client. The type of the ticket issued depends on the type of services the client is supposed to receive, and this is decided by the server.

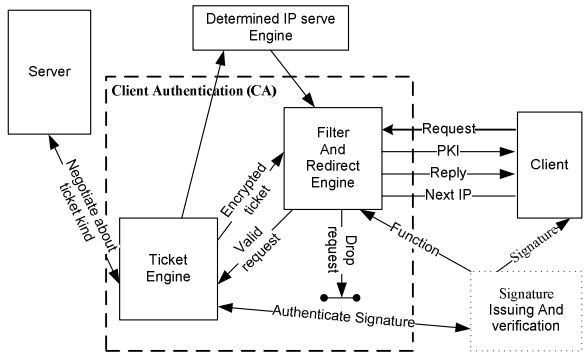

Fig. 3. HACSS Architecture

The CA contains two main parts: Filter and Redirect Engine and Ticket Engine. The Filter and Redirect Engine filters clients' requests, and the Ticket Engine will be responsible for the Clients' authentication. This Engine also communicates with Signature Issuing and Verification (SIV), a third party government organisation who is responsible for issuing signatures. In the following section, each part of this component will be explained.

### 3.2.1 Filter and Redirect Engine

Filter and Redirect Engine is the main window for the system to communicate with unauthenticated clients, and it provides a secure channel during their authentication stage. It has a minimum number of functions. This helps to prevent flooding and minimize the possibility of receiving malicious packets. Filter and Redirect Engine is responsible for the following tasks (Fig. 4):

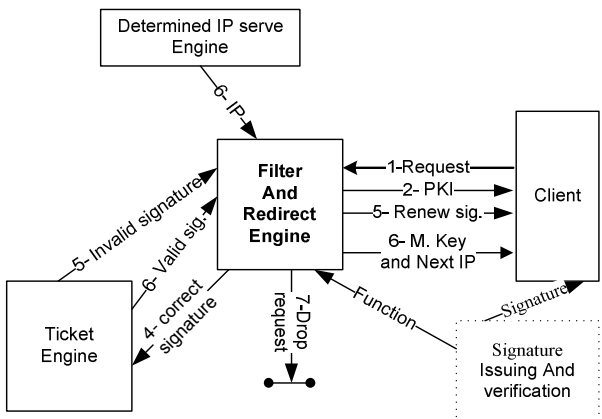

Fig. 4. Filter and Redirect Engine

1. It receives all unauthenticated clients' requests. In order to protect this area from malicious packet, these requests must be in a specific format and size and unencrypted. Any request does not match these criteria will be dropped. In addition, it processes only one request for a source in a specific time to stop flooding from the source. However this duration of time is changed dynamically, depending on the number of clients. This helps to prevent a client from engaging in the system for long time when this client's IP is spoofed. This function will be discussed later in this paper.

2. It sends its public key to the client after a correct client's request is received. This key is useful because the client's signature must be encrypted using this key when the signature is sent to the system. This will prevent sniffing and spoofing for that signature.

3. Each signature is checked by a function which is provided by the SIV. This function is used to check the signature's form and format. However this function is not responsible for validating signatures. The Filter and Redirect Engine receives this function from SIV through a secure channel.

4. It sends new correct signatures to the Ticket Engine to authenticate them.

5. When it receives a new invalid signature, it adds this signature to a black list which contains invalid signatures, and then it will request the client to renew its signature. Any received signature must be checked against this list before it is processed. This list consists of two columns for invalid signatures and numbers of times each invalid signature has been spoofed (Fig. 5). This list is sorted by the second column according to the frequency. Though these signatures are correct in format, they are no longer valid at the SIV. Additions of new signatures to the list take place after the signatures were rejected by the SIV. Although the checks against this list might be done partially, this still drop invalid signatures that might be used many times by attackers, and this also prevents other parts of the system from being occupied by excessive workload caused by these invalid signatures.

6. When a client is authenticated, the Filter and Redirect Engine sends the master key (encrypted by the client's PKI) to this client for dynamic key encryption. It also sends the next IP which this authenticated client can use to continue communication. This IP was determined by the Determine IP Serve Engine.

| Signature | No of Times | |
|-----------|-------------|---|
| Sign. 5 | 456 | |
| Sign. 101 | 430 | Part of the list |
| Sign. 6 | 399 | will be checked |
| .... | .... | |
| Sign. n | 1 | |

Fig. 5. Black Signatures List

7. It drops any request from clients who are authenticating or have been authenticated. So it stops flooding that is caused by spoofed and reused packets of those clients. In addition, it also drops any request which is not in the specified format. Because it only accepts one request at a time from a source, it drops all following packet from that source during that time. It also drops incorrect signatures and any received sig-natures which exist in the black list of invalid signatures.

The HACSS adapts the dynamic key technique to enhance our approach to stop DoS attacks by preventing packet sniffing. A dynamic key is a private key that is based on one time password technique. This technique will change the key for every packet. So it will be hard for the attacker to crack the keys [11], [12], [13].

### 3.2.2 Ticket Engine

The Ticket Engine authenticates clients' signatures and issues tickets for them (Fig. 6). This takes place through several steps. (1) It only receives valid clients' requests with their correct signatures. (2) It validates each client's signature by communicating with the SIV. (3) If the signature was invalid, it requests the client to renew his/her signature through the Filter and Redirect Engine, and add the invalid signature to the black list of invalid signatures in the Filter and Redirect Engine. (4) When the client's signature is valid, the Ticket Engine negotiates with the server about the details of a ticket with which this client can be granted. (5) After issuing the ticket for the client, a notification will be sent to the Filter and Redirect Engine that the client has been authenticated. Also it will send the authenticated signature to the Determine IP Serve Engine, which is in the ACC, to determine an IP with which this client should continue communication.

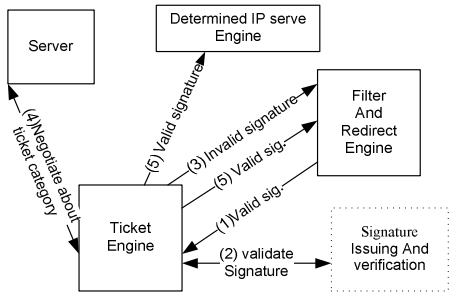

Fig. 6: Ticket Engine

### 3.3 Authenticated Client Communication (ACC)

This is the second component of the HACSS. It only communicates with authenticated clients via a full dynamic key encryption channel. In addition, it uses dynamic-multi-points-communication. This mechanism seeks an appropriate communication for each client by dynamically changing between multiple IPs. Each packet that arrives to the system must come through a specific IP. The ACC designs clients' packets, checks received packets, and checks whole clients' messages and files before they are sent to the server. In the following sections, each part of this component will be explained. The roles of each part of this component in the system are illustrated in the figure (Fig. 7).

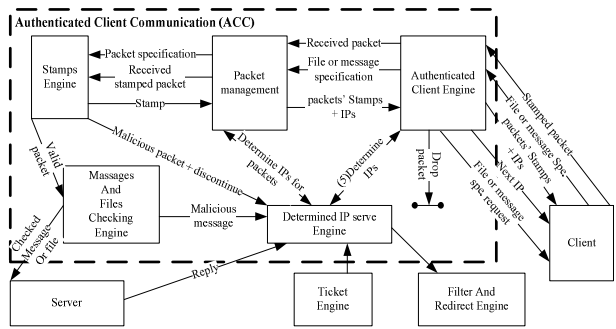

Fig. 7. Authenticated Client Communication (ACC)

### 3.3.1 Authenticated Client Engine

This engine consists of more than one IP. Each IP handles only specific packets and requests from the specific clients. These specific packets and requests are determined by the Determine IP Serve Engine. Every unexpected packet will be dropped at this stage. After each client's communication with the system, the client should receive a next IP that can be used to continue further communication. This mechanism provides a layer of packets filtering from their headers. In addition, this represents a filter layer for the malicious packet address attacks (Fig. 8).

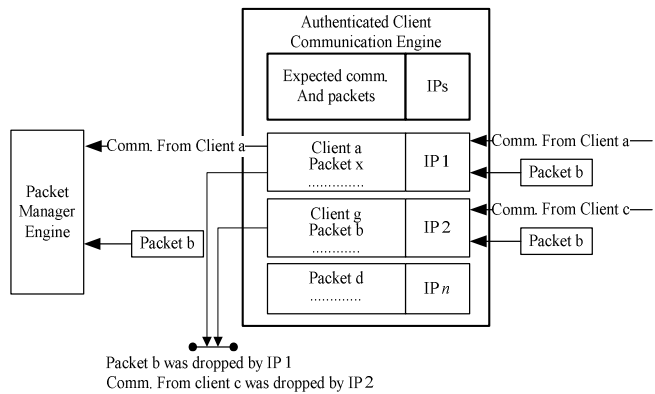

Fig. 8. dynamic-multi-points-communication

In Fig. 8, $IP_1$ received a communication packet from *Client a* and *Packet b*. The communication was accepted and was passed to the Packet Manager Engine, because it had been accepted by $IP_1$. However the *Packet b* was dropped because it had not been expected in this IP. On the other hand, in $IP_2$ the communication from *Client c* was dropped and *Packet b* was accepted, because the *Packet b* had been expected in this IP and communication from the Client c was not.

The communication between the client and the ACC is fully encrypted using dynamic key mechanism. To prevent flooding attacks, the ACC performs several steps in its communication with clients (Fig. 9). (1) When a client wants to pass a message or upload a file to the server, it is requested to provide specifications of the message or the file, such as size, file name, etc. (2) When these specifications are received, they will be passed to Packet Manager. (3) Authenticated Client Engine receives packets' stamps and their IPs from the Packet Manager, and then sends them to the clients. (4) It receives packets only from the clients that have stamps and their details attached to the encrypted part of the packet, as shown below in the figure (Fig. 10). (5) It receives a notification from Determine IP Serve Engine about clients' requests and packet specifications that the Authenticated Client Engine should receive for each IP. (6) It sends the next IP with which the client can continue further communication. (7) It drops any packet which has not been expected by the receiver IP, so the Authenticated Client Engine drops spoofed packets after they were decrypted.

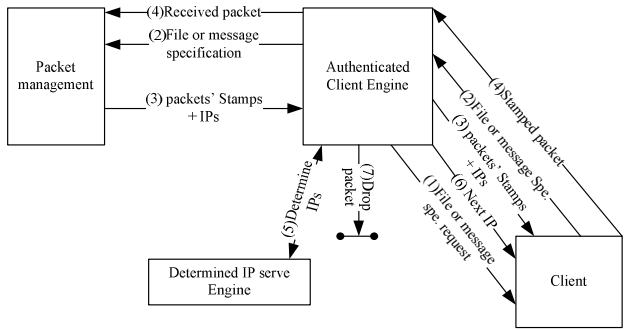

Fig. 9. Authenticated Client Engine

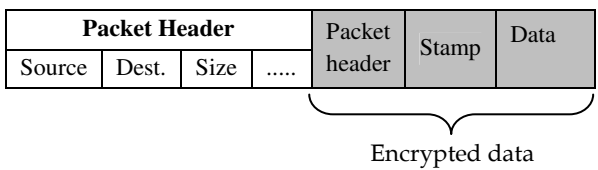

Fig. 10. Client's Packet

### 3.3.2 Packet Manager:

It is one on the important parts in the system that designed for the HACSS to stopping malicious packet attacks. It designs packets for the authenticated clients, so the system should not process any unexpected packet from them. It is responsible for the following tasks (fig. 11).

1. It receives specifications of each message or file to determine the number of packets that are required for the client to upload files or pass messages to the server.
2. It also determines the attributes for each packet which can be used as packet headers.
3. Each determined packet's header is sent to the Stamp Engine in order to issue a stamp for the packet.
4. It receives encrypted stamps for packets from the Stamp Engine.
5. The Packet Manager also sends each packet header to the Determine IP Serve Engine to specify from which IP this packet should be received.
6. It receives determined IP from the Determine IP Serve Engine for each packet.
7. The Packet Manager then sends all the appropriately determined packet headers, stamps and IPs through to the client.
8. When the packet has been received from the Authenticated Client Engine, it will be checked by the Packet Manager, so the Packet Manager is responsible for stopping duplicated packets.
9. It sends received packet to the Stamp Engine.

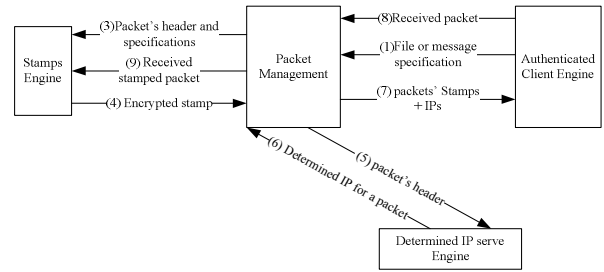

Fig. 11. Packet Manager Engine

### 3.3.3 Stamp Engine:

Stamp Engine is responsible for issuing stamps for packets and verifying stamps of the packets. Any authenticated client packets will be dropped unless they have specific stamps. Each stamp is issued by the Stamp Engine must have following characteristics:
- Stamps should represent the packet's attributes. Which are encrypted by the Stamp Engine.
- Each stamp should be different from others.
- Each stamp must represent the packet.
- A stamp must show its issuing date and time in case of stamps were reused.
- Each stamp must be encrypted so it is not readable.

Stamping packets technique is designed for the HACSS in order to stop malicious packet attacks. It is the main malicious packet filtering for authenticated clients' packets. The Stamp Engine performs the following tasks (Fig. 12):

1. It receives packet's header from the Packet Manager Engine.
2. It issues an encrypted stamp for the packet, and then sends this stamp to the Packet Manager Engine.
3. It receives clients stamped packets which have stamps from the Packet Manager Engine.
4. Once a packet is received, its stamp will be decrypted by the Stamp Engine.
5. Then the Stamp Engine will check the stamp's issuing date and time. If the details were satisfactory, then Stamp Engine will verify if this stamp represents its packet.
6. When the stamp represents its packet, then this packet will be moved to Massages and File Checking Engine.

If the stamp's issuing date and time are not correct, or the stamp does not represent the packet, the packet could be a malicious packet. The communication with that client then must be stopped, because this client is trying to perform a denial of service to the system.

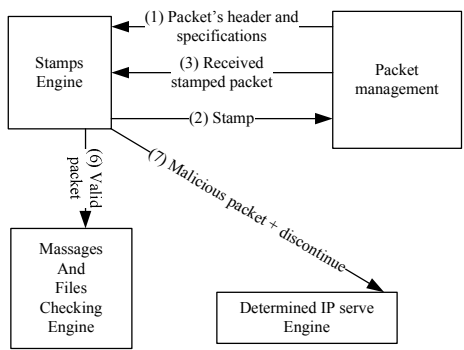

Fig. 12. Stamp Engine

### 3.3.4 Determine IP Serve Engine:

Determine IP Serve Engine determines the next IP with which the client should communicate after the client has been authenticated, and it also determines appropriate IPs for authenticated clients while communicating with the Authenticated Client Engine. Also it determines to which IP each packet should be sent. Each IP should receive a notification from the Determine IP Serve Engine about which clients should be served, and also which packets should be received by. In addition, it also maintains load balance between these IPs.

### 3.3.5 Messages and Files Checking Engine:

This component can be a virtual memory or separated device. It receives client packets, collects them and checks the whole message or file. It has a combination of security checking tools and programs like ant-viruses. After a message or file has been checked, it is sent to the server. When a file or a message has a malicious code or another threat, it will be dropped and then stop communication with the client.

## 4 DISCUSSIONS ON SECURITY AND EFFICIENCY OF THE HACSS

This section shows how the HACSS helps to prevent flooding attacks and to stop malicious packet attacks. Also the proposed approach and the existing methods will be compared. In addition, an efficiency analysis of implementing the HACSS will be conducted to evaluate the system in terms of flooding prevention and of malicious packet stopping.

### 4.1 Security Analysis

Clients can communicate with the system through two areas, namely the Filter and Redirect Engine and the Authenticated Client Engine. The following two sections will discuss the efficiency of these two areas in the proposed approach for both flooding and malicious packets attacks.

### 4.1.1 Preventing flooding attacks

In this section, the HACSS in preventing flooding for the Filter and Redirect Engine and the Authenticated Client Engine will be discussed. Also the proposed approach and the flooding prevention existing techniques will be compared.

#### 4.1.1.1 Filter and Redirect Engine

In this part of the system, the HACSS is designed to prevent flooding by implementing different techniques to address the following:

1. The HACSS minimises the number of system processes for a client's request in order to process the same quantity of requests using less the existing resources. So it can handle a more number of requests at a time. So the system will be harder to attack by flooding. This will be discussed in the Efficiency Analysis section (4.2).
2. The Filter and Redirect Engine processes only one request for a source in a specific time. This decision makes the system to prevent flooding. Usually the flooding attacker repeats sending a flooding packet which is generated with its machine protocol. So the source for these packets will be the same. Then the flooding from that source will be stopped. But in the case that the flooding attacker changes the source of these packets manually or using a program, the system should be faster to drop these packets, because generating packets manually or with such programs should significantly reduce the number of packets sent to the system. However, the duration of time to stop a specific source is changed dynamically, depending on the number of clients who are served on that time. This benefits the system and the client by preventing the clients from connection in case this client IP is spoofed in the system. This will also be discussed in the Efficiency Analysis section (4.2).
3. Usually flooding comes from unauthenticated clients. Once the client is authenticated, it will be moved to the Authenticated Client Engine. So more processes will then be made free to prevent flooding in this area.
4. The Filter and Redirect Engine has a function which is provided from the SIV. This function can be used to know the correct formats of the signatures. This process should be simple and focus on the signatures' formats to find incorrect signatures by using less process. This should protect the system from losing time to handle incorrect signatures.
5. The Filter and Redirect Engine has a black list containing invalid signatures. This list prevents the system from spending time to communicate with the SIV to verify invalid signatures. Also the sorting in this list saves time to find invalid signatures that were frequently used.

These pervious techniques are incorporated together to save time for the system to find incorrect requests for which the system should not spend much time. This saved time becomes important when more flooding attacks are received. So the availability of the system should be higher and the system should be less likely vulnerable against flooding attacks.

### 4.1.1.2 Authenticated Client Engine

In the Authenticated Client Engine, the dropping of flooding is easier, because in the dynamic-multi-points-communication, the packet headers prescribe the action to be taken. Also the Authenticated Client Engine address the following points:

- Because each IP already knows from which source the request or the packet should come, the Authenticated Client Engine drops any received request or packet that is not expected to be received. This technique saves time of processing flooding attacks, and this saved time will be helpful when more quantity of flooding occurs. This will be discussed further in the Evaluation Efficiency section for possible implementation.
- In the case a packet is spoofed and sent to the correct IP, the packet will be dropped after it decrypted, because the source and destination in the encrypted part will be incorrect.

The active queue management policy, which was mentioned in the previous section (2), increases the size of the backlog queue to offer more free space for new requests to prevent flooding. Also the RD flooding solution drops queued requests randomly to create more space for new requests. The limitation in these techniques is that the system is still limited by its resources. So actually this solution just uses the maximum existing resources to improve the system availability. However the HACSS changes the functionality of the system so it can serve more requests at a time within the same system resources. This happens, as discussed above, by offering a solution to drop adversary flooding attacks with less process. So, more availability of the system's resources is saved in case more flooding occurs in future. So the HACESS improve the availability of the system by enhancing the capability of the same system's resources to handle more quantity of flooding.

#### 4.1.2 Malicious packet attacks

In the Filter and Redirect Engine, most communication services are stopped. Only the communication services for client authentication run in this part of the system. This is particularly effective because required many kinds of malicious packet attacks cannot be committed unless the system's vulnerable services are running. Limiting the services thus reduces the risk of the system being attacked by malicious packets attacks in this area. In addition, all clients' requests and communication in this area are limited and must be in specific formats and size, otherwise they will be dropped before they get into the system.

On the other hand, all required communication services are running in the Authenticated Client Engine. Each client can use any one of these services depending on his/her ticket. Usually the attacker commits the malicious packet attack by sending a packet with changed attributes. When the system protocol was vulnerable for this change then a malicious packet attack would occur. Using stamp mechanism which was designed for the HACSS, the acceptance of this change can be avoided, because the system assigns these attributes and ensure that they have not been changed. Also though checking these attributes is done for each packet before collecting packets of each message or file, because some malicious packets have correct attributes for their packet levels, but it might be vulnerable for the system when the system tries to join them together.

Usually, malicious packet attacks are committed by unauthenticated clients. This kind of clients cannot communicate with the system through this part, unless an authenticated client packet was spoofed. If so, the chance of this packet getting into the system is low, because the Authenticated Client Engine uses the dynamic-multi-points-communication technique. If this packet was expected by the receiver IP, then it will be dropped after decrypted, because the header attributes will be included in the encrypted part. If the dynamic key encryption was encrypted, this malicious packet will be dropped by the Stamp Engine because its stamp will be incorrect.

From the above analysing, we can see that the HACSS should stop all current and future malicious packets, especially in the Authenticated Client Engine, because the client's packets are designed by the server side. If the authenticated or unauthenticated client tries to send a malicious packet to the server, this packet will be found malicious from its stamp.

In the following table (Table 1), the two previous solutions, Decision Tree and Deep Network Packet Filtering, mentioned in the section 2 will be compared with the HACSS.

TABLE 1
COMPARING THE TWO TECHNIQUES OF STOPPING MALICIOUS PACKET ATTACKS WITH HACSS

|  | Defence against existing attacks | Defence against new attacks | Guarantee all correct packet will be passed |
|---|---|---|---|
| Decision Tree | YES | NOT SURE | NO |
| Deep Network Packet Filtering | YES | NO | YES |
| HACSS | YES | YES | YES |

Table 1 shows that all the three solutions against malicious packets attacks can stop all the existing malicious packet attacks. However, the Deep Network Packet Filtering cannot stop new malicious packet attacks, because it was designed to find the existing malicious packet attacks. The Decision Tree was designed to find new malicious packet attacks by using learning algorithms to take a decision about packets. However, this algorithm does not guarantee to stop all future malicious packet attacks. The HACSS on the other hand should stop all future malicious packets, because most of the client's packets are designed and checked by the server side. The Deep Network Packet Filtering and the HACSS should guarantee that all non-malicious packets will be passed. On the other hand, the Decision Tree might drop non-malicious packets, because its learning algorithms might take incorrect decision. From Table 1, we can say that the HACSS is an enhanced solution to stop malicious packet attacks.

## 4.2 Efficiency Analysis

In this section, the performance of the HACSS to prevent flooding and to stop malicious packet will be discussed.

### 4.2.1 Flooding prevention

This discussion will focus on how the HACSS effectively saves time while preventing flooding. The discussion will be for both system's windows: Filter and Redirect Engine and Authenticated Client Engine.

#### 4.2.1.1 Filter and Redirect Engine

The HACSS minimises the number of processes for a request to increase the number of clients that can be served at a time. This can be verified in the following equation:

$$n = \frac{s}{p} , \qquad (1)$$

Where $p$ is the number of processes for each single client's request, $s$ is the number of processes can system handles at a time.

As seen in (1), a smaller number of processes for each client's request would result in a large number of clients who can be served at a time $n$.

As we mentioned before, HACSS only accept a single request from a source at a time. The HACSS makes the duration of time to stop the source changing dynamically to prevent a client from communicating with the system for long time, in case this client source is used in a flooding attack. This dynamic time is determined depending on the percentage of the system's usage. It can be calculated as follows:

$$D = t \times \frac{c}{n} , \qquad (2)$$

Where $t$ is the fixed blocking time for a source, $c$ is the number of clients that are serving at the time, $n$ is the number of clients that the system can serve at a time.

In (2), when the number of clients who are served at a time is small, the dynamic time for blocking multiple requests $D$ will be reduced. So when the system has an ability to receive more requests, this dynamic time is reduced to give the client a chance to reach the system.

#### 4.2.1.2 Authenticated Client Engine

In the Authenticated Client Engine, the dynamic-multi-points-communication mechanism is used to dropping flooding packets in less time. In the (3), the number of clients that the system can serve at a time $N$ consists of all clients' requests from both clients and adversary attackers.

$$N = N_c + N_f , \qquad (3)$$

Where $N_c$ is the number of clients' requests which the system can serve at a time, $N_f$ is the number of adversary attackers requests (from flooding attacks) which the system can serve at a time.

Also number of clients that can be served at a time ($N$) can be defined depending on the system processing divided by the number of processes for each request, as follows:

$$N = \frac{P}{P_{requests}} , \qquad (4)$$

Where $P$ is the number of processes that the system can do at a time, $P_{requests}$ is the number of processes that system does for each request.

So the number of processes $P$ that the system can do at a time ($P$) can be defined as follows:

$$P = (N_c * P_{requests}) + (N_f * P_{requests}) \qquad (5)$$

The time that the system needs to handle these processes can be defined, as in Eq. 6 below:

$$T = N_c * P_{c_{requests}} * t + N_f * P_{f_{requests}} * t , \qquad (6)$$

Where $P_{c_{requests}}$ is the average number of processes in each client's request, $P_{f_{requests}}$ is the average number of processes in each adversary attacker's request, $t$ is time required to handle each process.

In the HACSS, the communication between the client and the system is done through two system points: System and Redirect Engine and Authenticated Client Engine. In the rest of this section, we are going to investigate how the HACSS will respond to adversary attackers in the Authenticated Client Engine, and to analyse the time that the HACSS can save while preventing adversary attackers:

$$N' = N'_c + N'_f , \qquad (7)$$

Where $N'$ is the number of clients that the system can handle at a time, $N'_c$ is the number of client's right requests which the system can handle at a time, $N'_f$ is the number of adversary attackers' requests (from flooding attacks) which the system can handle at a time.

Because the implementation of the dynamic-multi-points-communication in this part of the system, the adversary attacker's packet might not be sent to a correct IP. A packet is sent to a correct IP when this packet's source is expected in the receiver IP. So, some of the adversary attackers' packets will be expected by the receiver IPs while the other will not. This can be calculated in the following equation:

$$N'_f = \frac{1}{I} N'_{f_{expected}} + \left(1 - \frac{1}{I}\right) N'_{f_{not\,expected}} , \qquad (8)$$

where $I$ is the number of IPs, $N'_{f_{expected}}$ is the number of adversary attackers' packets that are expected from the receiver IPs, $N'_{f_{not\,expected}}$ is the number of adversary attackers' packets that are not expected from the receiver IPs.

The time that the system needs to handle an adversary attacker's request $T'_{f_{request}}$ can be calculated as follows using (9):

$$T'_{f_{request}} = P'_{f_{requests}} * t , \qquad (9)$$

where $P'_{f_{requests}}$ is the average number of processes for each adversary attacker's request.

The time of handling adversary attackers' requests $T'_{f_{requests}}$ can be calculated using (10):

$$T'_{f_{requests}} = \left( \begin{array}{c} \left(\frac{1}{I} N'_{f_{expected}} \times P'_{f_{requests}}\right) + \\ \left(\left(1 - \frac{1}{I}\right) N'_{f_{not\,expected}} \times P'_{f_{requests}}\right) \end{array} \right) * t \qquad (10)$$

In the case the receiver IP is not expecting to receive the received packet, the packet will be dropped from its header, so $P_{f_{not\,expected}}$ will be equal to 0.

$$T'_{f_{request}} = \frac{N'_{f_{expected}} * t}{I} \qquad (11)$$

So the time of processing adversary attackers' requests or packets in the HACSS, $T''_f$ will be smaller as a result of $T'_{f_{request}}$ is reduced by $\frac{1}{I}$, as shown in (11) above.
$$N'' > N$$

Depending on the above evaluation, the system has more saved time that comes from finding and dropping flooding attacks. So the system capability is higher to prevent more quantity of flooding attacks.

### 4.2.2 malicious packet stopping

In this section, the performance of the HACSS to find malicious packets will be analysed. Duration of time that the system needs to identify malicious packets in the previous solution ($T'$) can be calculated as follows in (12):
$$T' = n \times t_c , \qquad (12)$$
where $n$ is the number of malicious packets, $t_c$ is the duration of time to check malicious packets.

In the HACSS the duration of time to identify the malicious packets (T'') can be calculated using the following equation (13):
$$T'' = n \times t_d , \qquad (13)$$
where $t_d$ is the duration of time to drop the malicious packets.

In the following, we are going to compare the difference between (T') and (T'') time:
$$T = T' - T'' \qquad (14)$$
$$\text{So} \quad T = n \times t_c - n \times t_d \qquad (15)$$
$$T = n(t_c - t_d) \qquad (16)$$

The time of dropping malicious packets is less than the time of checking packets, as below:
$$t_c > t_d \qquad (17)$$
$$T > 0 \qquad (18)$$

From the above evaluation, we can see that by using the HACSS, we save T time for checking malicious packets.

## 5 CONCLUSIONS AND FUTURE WORK

In this paper, we proposed a new security approach which we call the Holistic Approach for Critical System Security (HACSS). The HACSS is designed for critical systems like government systems, and it can be expanded to be implemented in other services systems. The new packet's stamp technique, the new dynamic-multi-points-communication mechanism, the division made in the client's communications areas, and the dynamic key encryption technique were included in the proposed solution. This comprehensive combination of these techniques makes the HACSS more powerful in preventing flooding attacks and stopping malicious packet attacks.

**Mohammed Alhabeeb** received the B.S (Hons) degree in Computer Information System from King Saud University in 1999 and M.S degree in network computing from Monash University in 2007. He is currently working towards the PhD degree at Monash University in Caulfield School of Information Technology. Mohammed's research interests include: denial of services, information security, and security analysis. Also, he is a project manager at the National Information Center, Ministry of Interior in Saudi Arabia.

Previous publications include:
- "Beyond Fixed Key Size: Classification toward a Balance between Security and Performance", AINA 2010: The International Conference on Advanced Information Networking and Applications, April 20-23 2010, Perth, Australia, In press.
- "Information Security Threat Classification Peyrmid", FINA: The Sixth International Symposium on Frontiers of Information Systems and Network Applications (FINA), April 20-23 2010, Perth, Australia, In press.

**Abdullah Almuhaideb** received the B.S (Hons) degree in Computer Information System from King Faisal University in 2003 and M.S degree in network computing from Monash University in 2007. He is currently working towards the PhD degree at Monash University in Caulfield School of Information Technology. Abdullah's research interests include: Ubiquitous Wireless Access, Mobile Security, Authentication & Identification. Also, he is a lecturer at the Computer Networks Department, King Faisal University.

Previous publications include:
- "Extended Abstract: an Adaptable Multi-Level Security based on different Algorithms Key Sizes for Mobile Devices", SIC 2009: the 3rd Saudi International Conference, June 5-6 2009, University of Surrey, Guildford, UK.
- "Comparative Efficiency and Implementation Issues of Itinerant Agent Language on Different Agent Platforms ", AT2AI-6: From Agent Theory to Agent Implementation Workshop in the scope of AAMAS 2008 ( The 7th International Conference on Autonomous Agents and Multiagent Systems), May 12-16 2008, Cascais, Portugal.

**Dr Phu Dung Le** is currently working at School of Information Technology. Dr Le's main research interests are: Image and Video Quality Measure and Compression, Intelligent Mobile Agents, Security in Quantum Computing Age. He used to teach Data Communication, Operating System, Computer Architecture, Information Retrieval and Unix Programming. He has also researched in Mobile Computing, Distributed Migration. Currently he is lecturing network security and advanced network security in addition to supervising PhD students.

Previous publications include:
- "A Tool for Migration to Support Resource and Load Sharing in Heterogeneous Environments", Proceedings of the International Conference on Networks, pp. 83-87, Feb 1996
- "A Limited-used Key Generation Scheme for Internet Transactions". Information Security Applications, Vol. 3325, pp 302-316, Lecture Notes in Computer Science, ISBN: 3-540-24015-2, Korea, 2005
- "The Design and Implementation of a Smart Phone Payment System", IEEE Proceedings of Information
  Technology: New Generations, pp. 458-463, USA 2006